\documentstyle[twoside,fleqn,npb,epsfig]{article}
\newcommand{\AmS}{{\protect\the\textfont2
  A\kern-.1667em\lower.5ex\hbox{M}\kern-.125emS}}

\title{Spin structure function of the virtual photon
 \thanks{Talk given by T. Uematsu at DIS 99, Zeuthen, April 1999}}
\author{Ken Sasaki\address{Department of Physics, Faculty of Engineering,
        Yokohama National University, \\ Yokohama 240-8501, Japan}
        and
        Tsuneo Uematsu\address{Department of Fundamental Sciences,
        FIHS, Kyoto University, \\
        Kyoto 606-8501, Japan}%
}

\begin{document}

\begin{abstract}
We investigate the spin structure of the virtual photon beyond the leading
order in QCD. The first moment of the virtual photon spin structure function
$g_1^\gamma(x,Q^2,P^2)$ with QCD effects turns out to be non-vanishing in 
contrast to the real photon case. Numerical analysis for virtual as well as 
real photon case is presented.
\end{abstract}

\maketitle

\begin{picture}(5,2)(-330,-330)
\put(2.3,-75){YNU-HEPTh-99-101}
\put(2.3,-90){KUCP-135}
\put(2.3,-105){May 1999}
\end{picture}

\section{INTRODUCTION}
In the last several years the nucleon's spin structure functions have been
extensively studied by deep inelastic scattering of polarized leptons on
polarized nucleon targets. 
Recently there has been growing interest in the polarized photon structure 
function. Especially its first moment has attracted much attention in the
literature in connection with the axial anomaly. 
Now the information on the spin structure of the photon
would be obtained by the resolved photon processes in the polarized electron
and proton collision in the polarized version of HERA.  More directly
the spin-dependent structure function of photon $g_1^\gamma$ can be measured
by the polarized $e^+ e^-$ collisions in the future linear colliders (Fig.1).

\begin{figure}[htb]
\hspace{1cm}
\includegraphics[width=5cm,height=5cm,keepaspectratio]{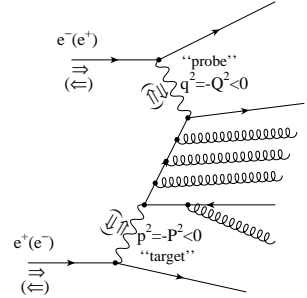}
\vspace{-1.8cm}
\caption{Deep inelastic scattering on a virtual photon in polarized
$e^+ e^-$ collision.}
\label{fig:polarized}
\end{figure}

We investigate the polarized virtual photon structure function 
$g_1^\gamma(x,Q^2,P^2)$ beyond the leading order in QCD, in the kinematical
region:
\begin{equation}
\Lambda^2 \ll P^2 \ll Q^2
\label{kin}
\end{equation}
where $-Q^2(-P^2)$ is the mass
squared of the probe (target) photon (Fig.1) with $\Lambda$ being the QCD
scale parameter. The next-to-leading order (NLO) analysis is now possible
since the required spin-dependent two-loop splitting functions in DGLAP 
evolution equation or equivalently the two-loop anomalous dimensions of
the relevant operators have been recently calculated \cite{MvN,V}.
The advantage in studying the virtual photon target is that we
can calculate the whole structure function by the perturbative method,
in contrast to the case of the real photon target, where there exist
non-perturbative pieces in NLO. Our motivation here is to carry out the
analysis of the polarized structure function at the same level as in the
unpolarized case \cite{UW}.

\section{QCD CALCULATION of $g_1^\gamma$}
The same framework used in the analysis of nucleon spin structure 
functions can be applied to the present case. Namely we can base 
our argument either on the operator product expansion (OPE) supplemented
by the renormalization group (RG), or on the DGLAP type parton evolution
equation.
The $n$-th moment of $g_1^\gamma$ for the kinematical region (\ref{kin})
is given by \cite{SU}
\begin{eqnarray}
&&\hspace{-0.5cm}\int_0^1 dx x^{n-1}g_1^\gamma(x,Q^2,P^2)=  
\frac{\alpha}{4\pi}\frac{1}{2\beta_0} \times \nonumber\\
&&\hspace{-0.5cm}\Bigl[
\sum_{i=+,-,NS}
L_i^n
\frac{4\pi}{\alpha_s(Q^2)}
\Bigl\{1-\left(\frac{\alpha_s(Q^2)}{\alpha_s(P^2)}\right)^{\lambda_i^n/2\beta_0
+1}\Bigr\}\nonumber\\
&&+\hspace{0.5cm}\sum_{i=+,-,NS}{\cal A}_i^n\Bigl\{1-\left(\frac{\alpha_s(Q^2)}
{\alpha_s(P^2)}\right)^{\lambda_i^n/2\beta_0}\Bigr\}\nonumber\\
&&+\hspace{0.5cm}\sum_{i=+,-,NS}{\cal B}_i^n\Bigl\{1-\left(\frac{\alpha_s(Q^2)}
{\alpha_s(P^2)}\right)^{\lambda_i^n/2\beta_0+1}\Bigr\}\nonumber\\
&&\vspace{2cm}\hspace{1.3cm}+\qquad {\cal C}^n +\quad {\cal O}(\alpha_s) \qquad
 \Bigr]
\label{master}
\end{eqnarray}
where $L_i^n$, ${\cal A}_i^n$, ${\cal B}_i^n$ and ${\cal C}^n$
are computed from the 
one- and two-loop anomalous dimensions together with one-loop
coefficient functions.
All of them are shown to be renormalization-scheme independent.
$\alpha_s(Q^2)$ is the QCD running coupling constant, and 
$\lambda_i^n \ (i=+,-,NS)$
denote the eigenvalues of one-loop anomalous dimension matrix.

\section{SUM RULE FOR $g_1^\gamma$}
For a real photon ($P^2=0$),  
the 1st moment vanishes to all orders of $\alpha_s(Q^2)$ in QCD
\cite{BBS}:
\begin{equation}
\int_0^1 dx g_1^\gamma(x,Q^2)=0.
\end{equation}
Now the question is what about the $n=1$ moment of the virtual photon case.
Taking $n\rightarrow 1$ limit of (\ref{master})
the first 3 terms vanish as
\begin{equation}
L_i^n \rightarrow 0, 
\sum_i {\cal A}_i^n\{ \ \} \rightarrow 0,
\sum_i {\cal B}_i^n\{ \ \} \rightarrow 0.
\end{equation}
Denoting $\langle e^4 \rangle=\sum_{i=1}^{n_f} e_i^4/n_f$ 
($e_i$: the $i$-th quark charge, $n_f$: the number of flavors), we have
\begin{equation}
{\cal C}^{n=1}=12\beta_0\langle e^4 \rangle (B_G^n+A_{qG}^n)|_{n=1}.
\end{equation}
Here we note that the sum of the one-loop coefficient function $B_G^n$
and the finite photon matrix element of quark operator $A_{qG}^n$ is 
renormalization-scheme independent and equal to $-2n_f$ for $n=1$.

Therefore we have
\begin{equation}
\int_0^1 dx g_1^\gamma(x,Q^2,P^2)=
-\frac{3\alpha}{\pi}\sum_{i=1}^{n_f}{e_i}^4
+{\cal O}(\alpha_s)
\end{equation}
We can go a step further to ${\cal O}(\alpha_s)$ QCD corrections
which turn out to be \cite{SU}
\begin{eqnarray}
&& \int_0^1dx g_1^\gamma(x,Q^2,P^2)\hspace{3cm}\nonumber\\
&&=-\frac{3\alpha}{\pi}
\left[\sum_{i=1}^{n_f}e_i^4\left(1-\frac{\alpha_s(Q^2)}{\pi}\right)
\right.\hspace{1cm}\nonumber\\
&&\hspace{1.0cm}-\left.\frac{2}{\beta_0}(\sum_{i=1}^{n_f}e_i^2)^2\left(
\frac{\alpha_s(P^2)}{\pi}-\frac{\alpha_s(Q^2)}{\pi}\right)\right]\nonumber\\
&&+{\cal O}(\alpha_s^2).\hspace{3cm}
\end{eqnarray}
This result coincides with the one obtained by Narison, Shore and 
Veneziano in ref.\cite{NSV}, apart from the overall sign for the definition
of $g_1^\gamma$.

\section{NUMERICAL ANALYSIS}

The polarized structure function $g_1^\gamma(x,Q^2,P^2)$ as a function of
$x$ is obtained by the inverse Mellin transform of the moments (\ref{master}).
In Fig.2, we have plotted the result for $n_f=3$, $Q^2=30$GeV$^2$, 
$P^2=1$GeV$^2$ with $\Lambda=0.2$GeV. Here we have shown
the Box (tree) diagram contribution:
\begin{equation}
g_1^{\gamma({\rm Box})}(x,Q^2,P^2)
=(2x-1)\frac{3\alpha}{\pi}n_f\langle e^4\rangle\ln\frac{Q^2}{P^2}
\end{equation}
and the Box diagram contribution including non-leading correction
(Box, NL),
the leading order (LO) QCD correction and the next-to-leading order (NLO)
QCD correction. In this analysis we observe that i) The NLO QCD correction 
is significant
at large $x$ as well as at low $x$. ii) No sizable change for the normalized
structure function is seen for different values of $Q^2$ and $P^2$. iii)
The $n_f=4$ case has been examined as well, but the normalized structure 
function is insensitive to the number of active flavors, $n_f$. 

\begin{figure}[htb]
\vspace{-2cm}
\includegraphics[width=8cm, height=10cm,keepaspectratio]{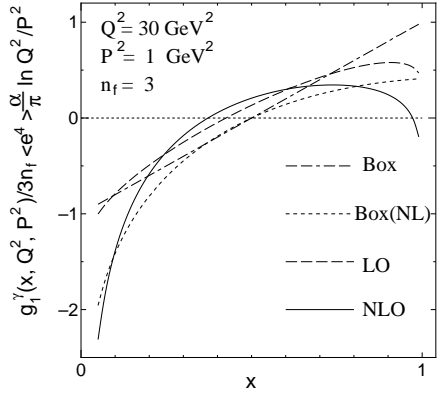}
\vspace{-2cm}
\caption{Spin structure function, 
$g_1^\gamma(x,Q^2,P^2)$ in units of $(3\alpha n_f\langle e^4\rangle/\pi)
\ln(Q^2/P^2)$, for $Q^2=$ 30GeV$^2$, $P^2= $1GeV$^2$, $n_f=3$ and $\Lambda=0.2$
GeV.}
\label{fig:virtual}
\end{figure}

\section{CONCLUDING REMARKS}
In this talk we have discussed the virtual photon's spin  structure function
$g_1^\gamma(x,Q^2,P^2)$ for the kinematical region (\ref{kin}), to the NLO
in QCD.
The result we obtained is independent of renormalization scheme.
The first moment of $g_1^\gamma$ is non-vanishing in contrast to the real 
photon case, where we have vanishing sum rule which is an extension of
Drell-Hearn-Gerasimov sum rule.
The NLO QCD corrections are significant at large $x$ as well as at low 
$x$.  The real photon case ($P^2=0$) consists of the perturbative piece and
non-perturbative piece, as 
\begin{equation}
g_1^\gamma(x,Q^2)=g_1^\gamma(x,Q^2)|_{\rm pert.}
+g_1^\gamma(x,Q^2)|_{\rm nonpert.}
\end{equation}
The perturbative parts can formally be recovered by setting
$P^2=\Lambda^2$ in (\ref{master}). In Fig.3 we have plotted the
point-like piece of $g_1^\gamma(x,Q^2)$ for $Q^2=30$GeV$^2$.
\begin{figure}[htb]
\vspace{-2cm}
\includegraphics[width=8cm, height=10cm,keepaspectratio]{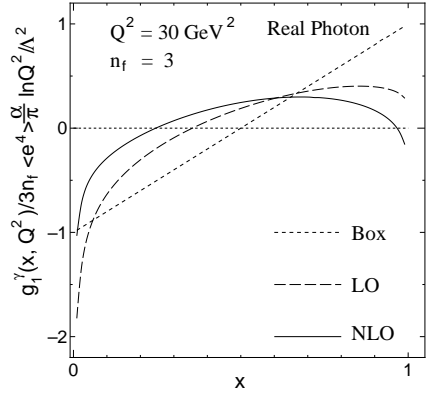}
\vspace{-2cm}
\caption{Real photon spin structure function $g_1^\gamma(x,Q^2)$ in units of 
$(3\alpha n_f\langle e^4\rangle/\pi)
\ln(Q^2/\Lambda^2)$, for $Q^2=$ 30GeV$^2$, $n_f=3$ and $\Lambda=0.2$
GeV.}
\label{fig:real}
\end{figure}
The LO results coincides with result by Sasaki in ref.\cite{S}, in which
the structure function $W_4^\gamma=g_1^\gamma/2$ was used.
The NLO results are qualitatively consistent with the analysis by
Stratmann and Vogelsang \cite{SV}.
The future subjects to be pursued are the following; i) We have to understand
how the transition occurs from the vanishing first moment for real photon 
($P^2=0$) to non-vanishing one for virtual photon ($P^2 \gg \Lambda^2$).
ii) We should study the spin-dependent parton distribution functions inside
the polarized virtual photon. The preliminary result shows that the NLO
effects are significant at small $x$ and also at large $x$ in the 
$\overline{\rm MS}$ scheme \cite{SU2}. 
iii) It would be intriguing to investigate another structure function 
$g_2^\gamma(x,Q^2,P^2)$ for which twist-2 and twist-3 operators contribute.
iv) The power corrections of the form $(P^2/Q^2)^k (k=1,2,\cdots)$ arising
from target mass effects and higher-twist effects should be explored. v)
An extension to time-like polarized virtual photon in $e^+ e^-$ process is
now under study.

\end{document}